\documentclass[aps,prd,nofootinbib,twocolumn]{revtex4}
\usepackage{amssymb}
\usepackage{amsmath,color}
\usepackage{epsfig}
\usepackage{graphicx,epsf,epsfig}
\usepackage{bbm}
\usepackage{subfigure}
\usepackage{multirow}
\usepackage{setspace}
\usepackage{verbatim}
\usepackage{epstopdf}
\usepackage[linktocpage]{hyperref}
\usepackage{multirow}
\newcommand{\bea}{\begin{eqnarray}}
\newcommand{\eea}{\end{eqnarray}}
\newcommand{\be}{\begin{equation}}
\newcommand{\ee}{\end{equation}}
\newcommand{\bp}{\begin{split}}
\newcommand{\ep}{\end{split}}

\begin{document}

\title{Examining the end of inflation with primordial black holes mass distribution and gravitational waves}

\author{Amjad Ashoorioon$^{1}$}
\email{amjad@ipm.ir}

\author{Abasalt Rostami$^{1}$}
\email{aba-rostami@ipm.ir}

\author{Javad T. Firouzjaee$^{2}$}
\email{firouzjaee@kntu.ac.ir}

\affiliation{$^1$School of Physics, Institute for Research in Fundamental Sciences (IPM), 	P.O. Box 19395-5531, Tehran, Iran}
\affiliation{$^2$Department of Physics, K. N. Toosi University of Technology,  P.O. Box  15875-4416, Tehran, Iran}

\begin{abstract}
We explicitly construct a double-field inflationary model, which satisfies the latest Planck constraints at the cosmic microwave background (CMB) scales and produces the whole dark matter energy density as primordial black holes (PBHs), in the mass range $10^{-17}~M_{\odot}\lesssim M_{{}_{\rm PBH}}\lesssim 10^{-13}~M_{\odot}$. The PBHs can be produced after the end of slow-roll inflation from the bubbles of true vacuum that nucleate during the course of inflation. Obtaining PBHs in this mass range enforces the scale of inflation to be extremely low, $10^{-7} \lesssim H \lesssim 10^{-3} ~{\rm GeV}$, which makes the efforts to observe gravitational waves at the CMB scales futile, although it is high enough to allow for a successful big bang nucleosynthesis. We will show that the shape of the mass distribution of the PBHs is dependent on how inflation ends and the Universe settles from the metastable direction to the true one. End of inflation can also be probed by examining the gravitational waves spectrum. In particular, we show that if exit from the rolling metastable direction to the true vacuum of the potential happens through a first-order phase transition after the end of slow-roll inflation, it leaves behind a stochastic gravitational wave background (SGWB), which is potentially observable by the Laser Interferometer Space Antenna. Examining the mass distribution of PBHs and possible SGWB from the end of inflation, we may be able to gain invaluable information about the end of inflation.
\end{abstract}

\maketitle

\section{Introduction}
\label{section:introduction}

In the absence of detection of weakly interacting massive particles (WIMPs) at the LHC and following the detection of gravitational waves by the Laser Interferometer Gravitational Wave Observatory (LIGO)  from the merging of the binary of black holes with masses of the order of a couple of decades heavier than the Sun \cite{TheLIGOScientific:2016src,Abbott:2016nmj}, the possibility of  black holes being candidates for dark matter has received an enormous amount of attention \cite{Bird:2016dcv,Sasaki:2016jop}. Although it is quite acceptable that these black holes are astrophysical, it will also be intriguing if they are primordial, namely, if they have been produced during the early Universe phases from large overdensities of radiation or matter. In particular, a recent LIGO/VIRGO merging event, GW190521 \cite{Abbott:2020tfl}, reveals  component masses that cannot be easily explained by the pair instability in the standard stellar structure theory ({\it cf.} \cite{Sakstein:2020axg} for some variations of the standard model that can fill this caveat; also, see \cite{Lehoucq:2009ge} for the accretion condition that needs to be satisfied, for the black holes to be primordial). This may further raise the stakes for primordial black holes (PBHs), being more than just figments of imagination.

In fact, since PBHs lighter than $10^{15}$ g could have evaporated by today, they have already been suggested to play a significant role in a variety of astrophysical phenomena, such as galactic \cite{Page:1976wx} and extragalactic \cite{Lehoucq:2009ge}  $\gamma$-ray background radiation or short-period bursts \cite{Cline:1996zg}, and the reionization of pregalactic medium \cite{Belotsky:2014twa}. On the other hand, heavier PBHs that are formed during the radiation dominated era are not subject to the constraints that baryonic matter is contingent on from big bang nucleosynthesis (BBN), about their contribution to the energy budget of the Universe and therefore, they act as nonbaryonic cold dark matter (CDM); please see \cite{Sasaki:2018dmp} and \cite{Carr:2016drx} for a comprehensive review.

Still, the contribution of the PBHs are very much constrained observationally and theoretically in various mass ranges \cite{Tisserand:2006zx,Nemiroff:2001bp,Wilkinson:2001vv,Carr:1997cn}. The only current open windows are in the sublunar mass range $10^{-17}-10^{-13}~ M_{\odot}$, in which the PBHs can constitute the whole dark matter energy density \cite{Sasaki:2018dmp}. The other open window is $10$-${\rm few}\times 100~ M_{\odot}$ PBHs in which the PBHs can make up to $10\%$ of CDM energy density. The interesting point about the first window is that  astrophysical black holes cannot occupy this mass range, as any star mass has to be at least about three solar masses \cite{Rhoades:1974fn} to collapse into a black hole. Thus any evidence for black holes in this mass region provides a massive boost to the idea that PBHs may have contributed significantly to the CDM energy density.

Since 1971 when Hawking raised the possibility that highly overdense inhomogeneities during the radiation dominated era can gravitationally collapse to a black hole \cite{Hawking:1971ei}, a host of scenarios have been put forward to explain the origin of PBHs in various mass ranges. Hawking himself suggested the possibility that PBHs may form from the collapse of cosmic string loops \cite{Hawking:1987bn,Polnarev:1988dh}. Such PBHs emit $\gamma$-rays, which their lack of observation puts an upper bound $G\mu\lesssim 10^{-6}$, which is only an order of magnitude larger than the one obtained from SGWBs \cite{Sanidas:2012ee} from the European Pulsar Timing Array and the PLANCK 2013 data \cite{Ade:2013xla}. Along with Moss and Stewart \cite{Hawking:1982ga}, he also showed how the collision of enough number of bubble walls during a first-order phase transition (PT) can lead to the formation of PBHs. This is one of the mechanisms that we will check to see if PBHs are formed in our scenario.

Inflation is a prime candidate for generating primordial fluctuations in the early Universe cosmology. These fluctuations later reenter during the radiation or matter-dominated phases and lead to the structures we observe today. In fact, the collapse of large density fluctuations is another mechanism of the formation of PBHs \cite{Carr:1974nx}. In this scenario, horizon reentry of a mode with large fluctuating amplitude during the ensuing radiation domination era leads to domination of gravity overpressure in the over-dense region and its collapse to the black hole. The amplitude of scalar fluctuations in the CMB scales is relatively small, $\mathcal{P}_{{}_S}\sim 2 \times 10^{-9}$ \cite{Akrami:2018odb}, and if the power spectrum is almost scale-invariant up to the scales relevant for PBH formation, their probability of formation is extremely small, $\sim \exp(-10^8)$. Therefore it is necessary that one would enhance the spectrum to at least $\mathcal{P}_{{}_S} \sim 10^{-1}$, at the relevant scales. Relating the mass of the PBHs to inflation provides us with exceptional information about the primordial power spectrum and inflationary potential, and other properties. Since with lensing, pulsar timing, and other astrophysical methods \cite{Schutz:2016khr,Monroy-Rodriguez:2014ula,Gaggero:2016dpq,Yokoyama:1998xd}, we can substantially constrain not only the initial energy density of PBHs $\beta(M_{{}_{\rm PBH}})=\rho_{{}_{\rm PBH}}/\rho$ in various mass ranges, but also the shape of the primordial power spectrum at the scales the CMB probes do not have any access (although, see \cite{Cole:2017gle} for the shortcomings). One other advantage of this approach to the other ones is that it can produce PBHs in the sub- or supersolar mass region, depending on the size of the horizon when the wavelength of the mode with large amplitude reenters.

Several mechanisms have been suggested along with this direction. For example,  two-field inflationary models with a large non-canonical kinetic coupling which induces two stages of inflation. In such models, at the transition between the two stages, the isocurvature perturbations temporary become  tachyonic and leads to an enhancement of the power spectrum \cite{Braglia:2020eai}. Another approach is flattening the potential and using the ultra slow-roll inflation to enhance the primordial power spectrum at the scales of interest for PBH formation  \cite{Garcia-Bellido:2017mdw,Germani:2017bcs}. Although this mechanism has been used, for example, in the context of Large Volume Scenario (LVS) inflation \cite{Cicoli:2018asa} to explain the origin of PBHs in the mass range  $10^{-17}-10^{-13}~  M_{\odot}$,  there are claims in the literature \cite{Motohashi:2017kbs} that it is difficult to go beyond $\mathcal{P}_{{}_S} \simeq 10^{-4}$ for the scales relevant within this mass range. This is not suitable for PBH formation during inflation \cite{Germani:2018jgr}, which even in the case of broad enhancement requires $\mathcal{P}_{{}_S}\simeq 10^{-2}$.

In \cite{Ashoorioon:2019xqc}, we proposed a mechanism of enhancement of the primordial power spectrum generated during inflation, which is based on a sixth-order polynomial dispersion relation, motivated in the context of extended effective field theory of inflation, (EEFToI) \cite{Ashoorioon:2018uey,Ashoorioon:2018ocr}. The mechanism was based on this observation that with the sixth order polynomial dispersion relation if the coefficient of the quartic term is negative and the ratio of quartic coefficient to the sixth order one squared is smaller than a threshold, the power spectrum of primordial fluctuations gets amplified. Since in the  (Extended) EFToI formalism, it is legitimate to assume that the couplings in the unitary gauge action are time-dependent, one can envisage the parameters such that the quartic coefficient becomes small and positive at the CMB scales but turns negative around the scales relevant for PBH formation. The effect would roughly take around two to three e-folds to show up in the power spectrum for the parameters we chose. It was possible to enhance the power spectrum to the threshold needed for PBH formation with the width, which could be controlled by how long this period of negative quartic coupling lasts. The UV cutoff of the theory can be adjusted to be well beyond the Hubble parameter, and the strong coupling can be avoided if the coefficients of the interacting operators are simply set to zero \cite{EEFToI-Interacting-NG}.

Nucleating cosmological defects such as cosmic strings and domain walls during inflation has been considered as one of the mechanisms of producing seeds for PBH formation \cite{Garriga:1992nm}.  Domain walls, whose size is smaller than a  critical radius and fall within the cosmological horizon early on, collapse due to their tension, forming PBHs in the postinflationary phase. The ones that are larger than that exponentially expand under their gravitational repulsion, and wormholes connecting the surrounding FRW universe with the baby Universes inside are created, which finally closes up in the light-crossing timescale. This will lead to the creation of two black holes at the two mouths of the wormholes \cite{Deng:2016vzb,Deng:2017uwc}. The resulting PBH mass spectrum has a wide distribution of the masses; however, the heavier ones are those created much earlier in the course of inflation. The mass function at such massive PBHs will get suppressed due to the exponential expansion during inflation and peaks for the ones whose mass is about the horizon mass at the end of inflation.

In this work, using the mechanism of \cite{Deng:2016vzb,Deng:2017uwc}\footnote{In \cite{Liu:2019lul}, the authors suggest a two-field inflationary model with super-Planckian field excursion, and hence Hubble parameter around $10^{-5} M_{P}$ . They then claim that the model can be tuned to acquire a considerable amount of PBHs in any mass range, due to the time-dependence of the Euclidean action. Apart from obscure evaluation of the Euclidean action in their model, the model is not able to  obtain  substantial PBHs with mass range of few hundred grams, which is typical of the grand unified theory (GUT) scale of inflation. Such PBHs are subject to evaporation by now. This is because we expect the PBHs of the collapsing bubbles to be distributed around $M_{bh}\sim \mathcal{O}(\frac{M_{\rm{P}}^2}{H})$.}, we try to construct an inflationary scenario, compatible with the latest Planck data at the CMB scales \cite{Akrami:2018odb}, such that it produces the whole dark matter energy density today through this mechanism. The mass fraction of PBHs, even though broad, is dominated by the ones that nucleate toward the end of inflation, whose mass is given by the horizon mass at the time. Suppose we want to account for the whole (or sizable) amount of dark matter energy density. In that case, we are left with two windows of sublunar, $10^{-17}-10^{-13}~M_{\odot}$, and intermediate-mass black holes, ${\rm few}-10^{5}~M_{\odot}$ \cite{Niikura:2017zjd, Montero-Camacho:2019jte}. These two mass scales correspond to an extremely small Hubble parameter, $10^{-7} \lesssim H \lesssim 10^{-3} ~{\rm GeV}$ in the former case, and $10^{-20} \lesssim H \lesssim 10^{-25} ~{\rm GeV}$. Assuming instant reheating at the end of inflation, one can estimate the reheating temperature, respectively, $10^{6} \lesssim T \lesssim 10^{8}~{\rm GeV}$ and $ 10^{-1} \lesssim T \lesssim 10^{-4}~{\rm GeV}$. In the first case, the reheating temperatures are large enough to allow for successful BBN. In the latter, if the reheating temperature is larger than few MeV, one can still construct a successful inflationary scenario. Nonetheless, this will require that reheating be extremely efficient. In this work, we focus mainly on constructing an inflationary model with Hubble parameter that may give us sublunar PBHs, although the latter is also possible to realize within our setup. In principle, our model is flexible to realize inflation at any scale from the GUT scale down to $\gtrsim (0.01~{\rm GeV})^4$.

If the string theory landscape \cite{Susskind:2003kw} exists, rolling and nucleating both seem to be quite commonplace in such a landscape (see \cite{Vafa:2005ui} for obstructions in realizing such a scenario). In fact it might be possible that it inflation has happened through chains of nucleations in the landscape \cite{Freese:2004vs,Chialva:2008zw,Ashoorioon:2010vw} (again see \cite{Ashoorioon:2008pj} for the roadblocks along the path). It is also possible that inflation has happened through a couple of rollings and nucleations. In this work, we consider a double-field inflationary scenario \cite{Linde:1990gz,Adams:1990ds} where inflation happens by slow-rolling. Exit from inflation and settling to the true vacuum, nonetheless, can happen through slow-roll violation, tachyonic waterfall instability, first-order PT, and sometimes a combination of the aforementioned possibilities. Namely, inflation may end via the slow-roll phase, but settling to the true vacuum happens afterward via a first or second-order PT. The scenario was analyzed in \cite{Ashoorioon:2015hya} using the setup of extended hybrid inflation \cite{Copeland:1994vg}. The model though is in conflict with the latest Planck bounds on the tensor-to-scalar ratio, $r\lesssim 0.065$, \cite{Akrami:2018odb} \footnote{One could use excited initial states for scalar fluctuations, to lower the predicted $r$, \cite{Ashoorioon:2013eia}.}. In \cite{Ashoorioon:2015hya} it was anticipated that this is just the characteristic of the potential chosen along the rolling direction, which was an $m^2\phi^2$ model. Here, we replace the quadratic rolling direction with a near-inflection-point inflation \cite{Allahverdi:2006iq,Allahverdi:2006we,Ashoorioon:2009wa}, which has the benefit of having a free energy scale, while simultaneously can convince the small tensor-to-scalar ratio and scalar spectral index compatible with the latest Planck results. The other benefit of the model is that field excursion in the model remains sub-Planckian, and one would not run into difficulties with super-Planckian displacements \cite{Baumann:2014nda}.

As rolling along the direction with near-inflection-point inflation happens, another vacuum along with the other direction forms. As soon as these two vacua become degenerate, bubbles of true vacuum begin to form. These bubbles will act as seeds for PBHs once they collapse in the postinflationary phase. The tuning to the true vacuum has to be much less than 1 in order for inflation to last long enough. It is also necessary to exclude the overproduction of PBHs whose energy density exceeds the dark matter's current energy density. As mentioned, there are different possibilities for the termination of inflation, and they affect the mass distribution of PBHs that can come out of the process. What really matters for bubbles to turn to PBHs, is that we have to exit inflation in a way to allow the bubbles to lose their kinetic energy, come to rest with respect to the parent FRW universe, and then collapse in the postinflationary decelerating phase. A waterfall second-order PT will provide the bubbles (including the ones whose sizes are larger than the Hubble radius at the true vacuum) with ample time to go through all these effects. On the other hand, concluding inflation with an abrupt first-order PT to settle to the true vacuum will destroy these bubbles as the first-order PT only completes by the collision of the bubble walls formed in the sea of false vacuum. There is an intermediate scenario in which, however, inflation finishes by the violation of the slow-roll, the inflaton still rolls along the false vacuum direction, but the expansion is no longer accelerating. Finally, the field reaches a point where the nucleation rate per unit four-volume to the true vacuum reaches the critical value at which the first-order PT to the true vacuum completes \cite{Coleman:1977py, Callan:1977pt}. The time elapsed from the end of inflation to the first-order PT's termination allows for the collapse of most of these bubbles. In addition, the first-order PT allows a SGWBs \cite{Hogan:1986qda,Kosowsky:1992rz, Kamionkowski:1993fg,Huber:2008hg} to be produced. For reasonable nucleation time, the peak frequency of the produced gravitational waves falls in the LISA sensitivity band \cite{Audley:2017drz}, which correlates a high-frequency gravitational wave signal, with specific characteristics from a first-order PT \cite{Kosowsky:1992rz, Kamionkowski:1993fg, Huber:2008hg}, with the PBHs in the sublunar mass range. We also show that a negligible number of PBHs is formed from the collision of the bubble walls and the first-order PT.

The paper is structured as follows: In Sec. \ref{dblfieldinf}, we introduce the setup of double-field inflation with near-inflection-point inflation as the rolling direction, exemplifying the scenarios discussed above. In Sec. \ref{massfractionpbhs}, we estimate the mass fraction of the PBHs produced from the bubbles collapse and the ones created from bubble collisions, showing although the former can be arranged to produce the whole or substantial part of CDM, the latter's contribution to PBHs mass function is negligible with a PT at such energy scales.  In Sec. \ref{SBGWsPT}, in the cases where settling to the true vacuum happens through a first-order PT, we compute the SGWB resulting from the collision of bubbles. We show that in particular for the one where PT occurs after the end of the slow-roll violation, the model is capable of producing SGWB in the reach of LISA or other high-frequency gravitational wave probes that scan that region of frequency. We finally conclude the paper in the last section.  $M_{{}_{\rm P}}=(8\pi G)^{-1/2}=1$ in this paper, except for the places where it is explicitly written.

\section{Double-field inflation\label{dblfieldinf}}

\label{dblfieldinf}
In this section, we will go through the basics of an inflationary double-field model, in which termination of inflation and settling to the true vacuum can happen through either a second-order tachyonic rolling or a first- order PT \cite{Coleman:1977py,Callan:1977pt}. First-order PT can either mark the end of inflation or happen sometime after the accelerating phase's termination with the slow-roll violation. We argue that, in the case where settling to the true vacuum happens via tachyonic second-order rolling or when the first-order PT happens after the termination of inflation, one can produce large enough PBHs to account for the dark matter energy density of the Universe. In the latter case, where PT
happens after the end of inflation, there will be a chance
to produce a SGWB from the collision of bubbles.

With the collision of true vacuum bubble walls and their  natural conversion to radiation, inflation termination becomes efficient. That is why in the old inflationary  scenario, Guth tried to invoke this mechanism with a  trapped metastable vacuum to produce the accelerating  phase needed to solve the problems of standard big bang  (SBB) cosmology \cite{Guth:1980zm}. However, it was soon realized that,  in order for inflation to last long enough to solve the SBB  problems, the Universe might never recover from the  accelerated phase if the rate of decay was as much as it  was necessary for a graceful exit from the accelerated  expansion \cite{Guth:1982pn}. The ''new'' inflationary scenario was  proposed, in which the scalar field was rolling slowly on  a relatively flat potential, which would become steeper  toward the end of inflation \cite{Linde:1981mu,Albrecht:1982wi}. The coupling of the  rolling field to the other species would either perturbatively \cite{Abbott:1982hn,Dolgov:1982th} or nonperturbatively \cite{Kofman:1994rk,Kofman:1997yn} reheat the Universe.
\begin{figure}[t]
	\includegraphics [width=3in] {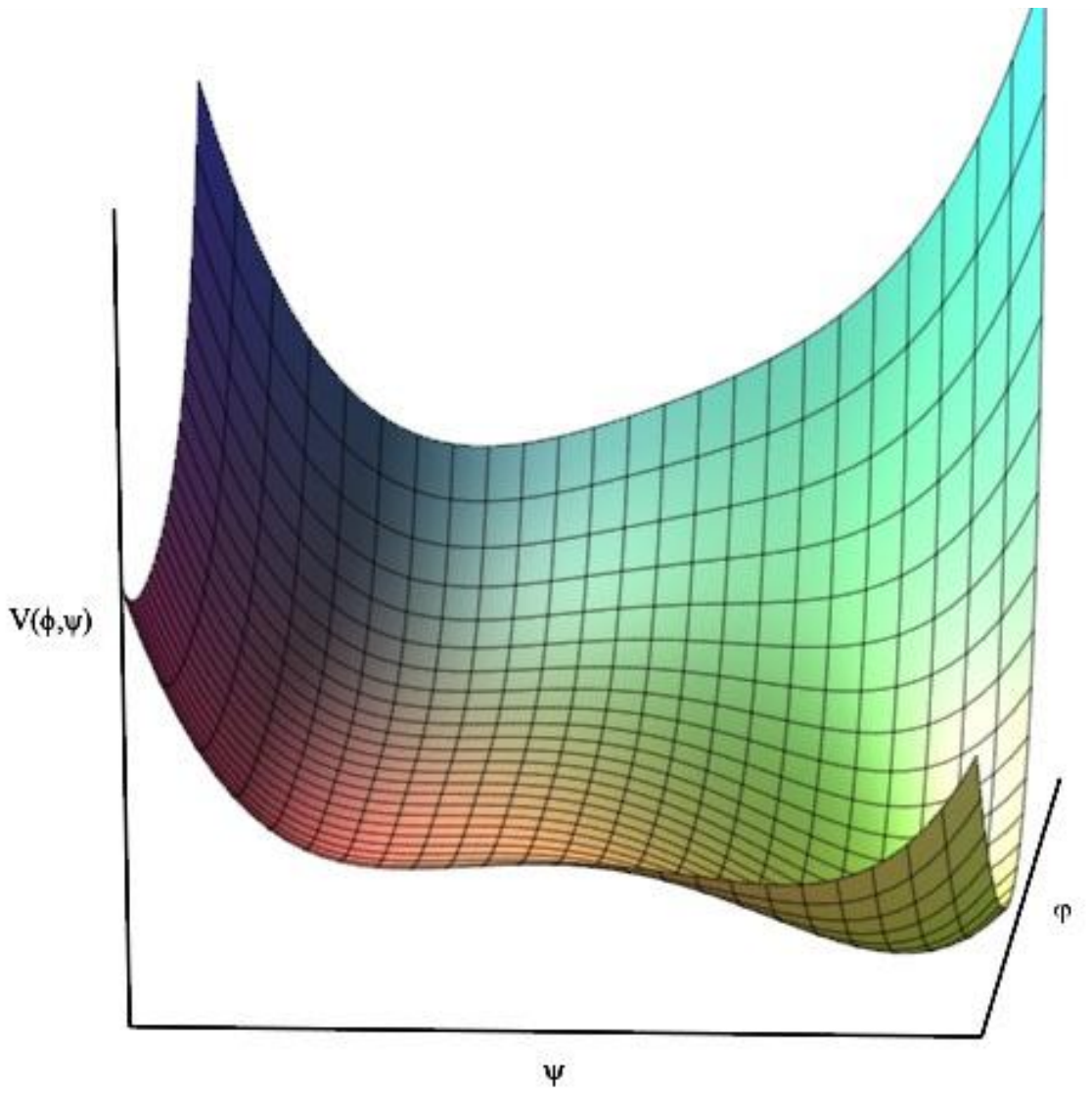}
	\caption{$V(\phi,\psi)$ with rolling inflaton $\phi$ and the trigger field $\psi$ that facilitates the PT or waterfall transition to the true vacuum. }
	\label{pot3d}
\end{figure}
Double-field inflation is a mixture of old and new inflationary scenarios \cite{Linde:1990gz, Adams:1990ds}, in the sense that one of the fields  rolls during inflation, as in the slow-roll inflation, and the  second field is initially trapped in the metastable vacuum.  (As can be seen in our example, this vacuum could be the  only existing minimum initially. The true vacuum could  develop as inflation proceeds.) As the first field rolls, either  the nucleation rate from the metastable vacuum to the true  one in the perpendicular direction grows and reaches the  critical value \cite{Guth:1982pn}
needed for the bubbles of a true vacuum to  percolate and cease inflation, or the second direction  becomes tachyonic and destabilizes the rolling field toward
the true minimum. The latter case is famously known as ''hybrid'' inflation  \cite{Linde:1993cn, Copeland:1994vg}. However, even before inflation ends this way, the bubbles of true vacuum are nucleated as soon as the two minima become degenerate, and they would collapse and lead to the formation of PBHs, as in \cite{Deng:2016vzb,Deng:2017uwc}. We will use the setup of double-field inflation to construct an explicit inflationary model that satisfies the latest Planck 2018 constraints at the CMB scales \cite{Akrami:2018odb} and generate PBHs with proper mass fraction after the end of inflation through the slow-roll violation. Following \cite{Ashoorioon:2015hya} \footnote{The model is in conflict with the latest Planck bounds on the tensor-to-scalar ratio, $r\lesssim 0.065$, \cite{Akrami:2018odb}. Nonetheless, one can use superexcited initial states for scalar fluctuations, to lower the predicted $r$ \cite{Ashoorioon:2013eia}.}, we assume the double-field potential takes the form 
\be\label{pot-tot}
V(\phi,\psi)=V_0+V_1(\phi)+ V_2(\phi,\psi),
\ee
where $V_0$ is the constant vacuum energy, and $\phi$ is the rolling field, whose potential $V_1(\phi)$, along with the vacuum energy, drives inflation. $\psi$ is the field, which facilitates settling down to the true vacuum either through tachyonic instability or first-order PT.

We focus on a variation of extended hybrid inflationary potential \cite{Copeland:1994vg,Ashoorioon:2015hya}, in which the potential along the rolling direction is replaced with an inflection-point inflation potential \cite{Allahverdi:2006iq,Allahverdi:2006we,Ashoorioon:2009wa}
\bea
&&V(\phi,\psi)=V_0+\frac{1}{2} m^2{\phi}^2 - \frac{ A \lambda_{{}_3}}{3} {\phi}^3 + \lambda_{{}_3}^2 {\phi}^4+  \nonumber\\
&&   \frac{1}{4}\lambda \psi^4-\frac{1}{3}\gamma M \psi^3 +\frac{1}{2} \lambda^{\prime} \phi^2\psi^2+\frac{1}{2} \alpha M^2 \psi^2
\label{Potential}
\eea
which is supplemented by the cubic term for the field $\psi$ to provide the possibility of tachyonic instability or first-order PTs at/after the end of inflation; please see Fig.(\ref{pot3d}). For large values of $\phi$, the potential has one minimum in both $\phi$ and $\psi$ direction. As inflation proceeds and $\phi$ rolls toward its vacuum, second minimum along the $\psi$ direction develops, if $\gamma^2> 4\alpha \lambda$ at
\be
\phi_{{}_{\rm I}}^2=M^2 \frac{\gamma^2-4\alpha \lambda}{4\lambda^{\prime}\lambda}\,.
\ee
When the value of $\psi$ drops below $\phi_{{}_{\rm I}}$, the potential $V_2$ acquires two minima, located at $\psi_{{}_{\rm{false}}}=0$ and $\psi_{{}_{\rm{true}}}$, which is given by
\begin{align*}
	\psi_{{}_{\rm{true}}} = \frac{2M\gamma + \sqrt{4M^2\gamma^2 - 36 M^2\alpha\lambda - 36 \lambda\lambda' \phi^2}}{2\lambda}\,.
\end{align*}
The two minima are separated by a barrier during inflation that allows for the nucleation of bubbles of true vacuum while inflaton rolls and can even allow for the termination of inflation with a first-order PT. The expression in the second line of q.(\ref{Potential}) plays the role of $V_2(\phi,\psi)$, which couples the inflaton to the field that creates the false vacuum in the $\psi$ direction and facilitates the phase PT. In principle, $V_1(\phi)$ determines the predictions of the model  at large scales in the limiting case that the cosmological constant goes to zero and can be chosen such that the model is compatible with the CMB observables at cosmological scales \cite{Akrami:2018odb}. We later add a constant to this direction to lift the inflating direction.

The model we are interested in exploits the inflection-point inflationary potential \cite{Allahverdi:2006iq,Allahverdi:2006we,Lyth:2006rze} for the rolling direction,
\bea\label{Hybrid-Pot}
V_1(\phi)&=& \frac{1}{2} m^2{\phi}^2 - \frac{A \lambda_3}{3} {\phi}^3 + \lambda_3^2 \phi^4\,.
\eea
This form of inflationary potential was initially proposed in the context of the potentials from supersymmetry breaking of the flat directions in the minimal supersymmetric standard model (MSSM), where the parameter $m$ is in the $1-10$ TeV energy scale. The $A$-term (like the mass term)  is a soft supersymmetry breaking parameter while the last one in (\ref{Hybrid-Pot}) comes from a minimal globally supersymmetric model. It was also realized in the context of Matrix inflation \cite{Ashoorioon:2009wa} that such an inflationary scenario is possible, although mainly the symmetry breaking potentials was focused on \cite{Ashoorioon:2011ki}. The advantage of the near-inflection-point potential is that while the scalar spectral index and the amplitude of the density perturbations at the CMB scales can match the observational data, the scale of inflation --and hence the tensor-to-scalar ratio $r$-- can be lowered as much as desired. The other benefit of this kind of inflationary potential is that the displacements of the field remain sub-Planckian. Here we assumed that such potential could be realized within the landscape of string theory \cite{Susskind:2003kw}. Regardless of the origin of such a potential, we assume that the scale of inflation is such that the Hubble parameter is in the range $10^{-7} \lesssim H \lesssim 10^{-3} ~{\rm GeV}$, which is desirable to produce PBHs as dark matter.

Theoretically, inflation can occur near a point $\phi_0$ if one sets the parameters such that $V'_1$ and $V'_2$ remain very small around this point. Indeed, choosing
\bea\label{exact-inflection}
\lambda_3 &=& \frac{m}{2\phi_0} \\ \nonumber
A &=& 4 m\,,
\eea
both the first and second derivatives of the potential exactly become zero at inflection point $\phi_0$.  For $\phi$ near the inflection point, $V'_1\approx 0$ and the background space-time is effectively de Sitter. However as time goes on,  the classical friction gradually leads to the slow-rolling evolution and eventually the inflationary phase ends with $\epsilon=\frac{1}{2}(\frac{V'_1}{V_1})^2\sim \mathcal{O}(1)$. Because of this extreme flat potential, the velocity of the rolling field is extremely small. Therefore, we expect that, in the low-energy scales, the potential (\ref{Hybrid-Pot}) with the parameters (\ref{exact-inflection}) gives rise to adequate amplitude for density perturbations explaining the Planck 2018 normalization \cite{Akrami:2018odb}. However, as discussed in \cite{Allahverdi:2006we}, for such inflection potential  the number of e-folds, $N_e$, of inflation after which the observable Universe leaves the horizon  is related to the spectral index $n_s$ as
\bea
n_{{}_S} \approx 1-4/N_e\,.
\eea
If the scale of inflation is about $10^8$ GeV and reheating is instantaneous and efficient, the required number of e-folds needed to solve the flatness and horizon problems of the SBB cosmology is  $N_e\approx 40$ \footnote{The required number of e-folds to solve the problems of the SBB for an inflationary model at the energy scale $M$ and reheating temperature $T_{\ast}$ is given by
	\bea
	N_e=53+\frac{2}{3}\ln(M/10^{14}{\rm GeV})+\frac{1}{3}\ln(T_{\ast}/10^{10}{\rm GeV})\,,
	\eea
	which for the inflationary scale of interest for us assuming instant reheating, yields $N_e\approx 40$.}, for which $n_{{}_S}=0.90$. Of course, this would have been an obstacle for such inflection-point inflation model. One way to obtain a larger scalar spectral index is to assume that one deviates from inflection-point condition such that  parameter $A$ is perturbed as
\bea\label{A-pert}
A = 4m\left(1-\frac{1}{4}\nu^2\right)^{\frac{1}{2}}\,.
\eea
Above, $\nu\ll 1$ is a dimensionless parameter that quantifies deviation from the inflection condition we had before. One can show that such potential still possesses an inflection point $\phi_i$ at which $V''_1(\phi_i)=0$ and up to quadratic order of $\nu$ is written as,
\bea
\phi_i = \phi_0 \Big(1 - \frac{1}{4}\nu^2 +\mathcal{O}(\nu^4)\Big)\,.
\eea
We should emphasize that due to the deviation of $A$, the condition $V'_1=0$ is not satisfied any longer at the new inflection point.  When inflation is driven near $\phi_i$, one can approximate the potential (\ref{Hybrid-Pot}) in the vicinity of the inflection point with its cubic Taylor expansion as\footnote{Comparing this expansion with that of \cite{Allahverdi:2006we}, the reader immediately recognizes that the coefficients of expansion are fairly different. This is because the coefficient of $\nu^2$ in the expansion of $\phi_i$ has been erroneously evaluated in \cite{Allahverdi:2006we}.}
\bea\label{inflection-point}
V_1(\phi)\approx\frac{1}{12} m^2\phi_0^2\Big(1&+&\nu^2\Big) + \frac{\nu^2}{4}m^2\phi_0\Big(\phi-\phi_i\Big)\\ \nonumber &+&\frac{2m^2-m^2\nu^2}{6\phi_0}\Big(\phi-\phi_i\Big)^3\,,
\eea
in which we have replaced $\lambda_3$ from (\ref{exact-inflection}) and $A$ from (\ref{A-pert}).

Apart from $V_0$, the parameters of inflection-point potential $V_1$ should be determined such that we retain the compatibility with the CMB observations at cosmological scales. Thus, for keeping the observational prediction extracted from $V_1$ intact, it is crucial to assume that $|V_0| \ll |V_1| $ during inflation. Moreover,  for having a zero vacuum constant after the PT at the minimum of total potential, which happens in, say $\phi=0$, we must impose the following relation between parameters of the potential $V_2$ and  $V_0$ term in (\ref{Hybrid-Pot}):
\be
V_2(\psi_{{}_{\rm{true}}},\phi=0)=-V_0\,.
\label{zero-constant}
\ee
Below we have taken $V_0\simeq \frac{V_1}{100}$. This will leave the CMB predictions derived for $V_1$ intact to $\mathcal{O}(10^{-4})$.

In the classical field theory, when the potential of a scalar field possesses two local minima $\phi_{\pm}$, for which $V(\phi_+) > V(\phi_{-})$, the vacuum at $\phi_{-}$ would be the unique classical vacuum of lowest energy. This minimum also defines the quantum theory's unique vacuum state (true vacuum) in perturbation theory. The classical vacuum $\phi_+$, known as false vacuum, is rendered unstable by barrier penetration due to the quantum fluctuations.  These quantum fluctuations cause bubbles of true vacuum to materialize in the sea of the false one, causing it to decay if the bubbles coalesce. When the bubble nucleation rate per unite four-volume $\Gamma$ becomes significant, the system undergoes a first-order PT due to bubble collisions and percolation.
\begin{figure}[t]
	\includegraphics [width=3in]{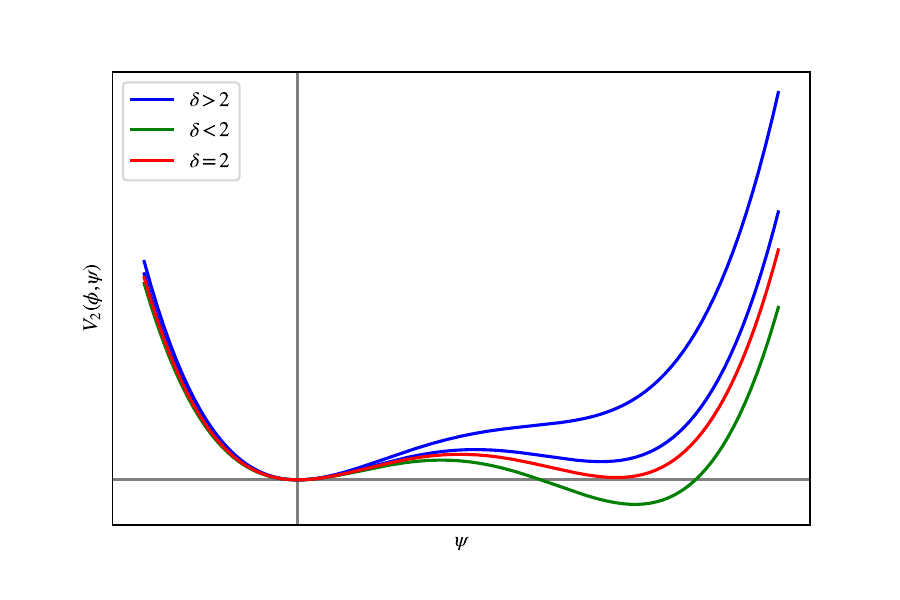}
	\caption{Evolution of $V_2(\phi,\psi)$ with rolling inflaton. The first blue line shows the $\phi={\rm const.}$ profile, when the potential attains its new minimum at $\phi=\phi_{{}_{\rm I}}$. The second blue curve illustrates the profile  when the field $\phi$ rolls further and the newly formed second minimum falls more. The red curve shows the profile, for $\phi_{{\rm critical}}$ when the two minimum become degenerate and $\delta=2$ . Finally the green curve shows the profile at $\phi < \phi_{{}_{\rm critical}}$, when the second minimum drops below the first one and  quantum tunneling from the first one to the true one acquires a nonzero probability. }
	\label{psi-direction-potential}
\end{figure}
According to \cite{Callan:1977pt}, one can derive the following relation for the bubble nucleation rate at the leading order  using WKB approximation:
\be
\Gamma= {\cal A} \exp(-S_{{}_{\rm E}})\,,
\ee
where $S_{{}_{\rm E}}$ is the on-shell four-dimensional Euclidean action satisfying the required boundary conditions of solution at the two minima. The prefactor  ${\mathcal A}$ with mass dimension four can be calculated using the path integral techniques, and so depending on the action, it may be completely difficult to elaborate explicitly \cite{Callan:1977pt}. Nonetheless  \cite{Linde:1981zj} suggests that the preexponential factor may be approximated as $M^4_{\psi}$ when $M^2_{{}_\psi} > 2 H^2$. Furthermore, it can be shown that, when $M^2_ {{}_\psi}< 2 H^2$, the magnitude of this prefactor is of the order of $H^4$ \cite{Linde:2005ht} \footnote{In \cite{Ashoorioon:2015hya}, it was realized that the correct estimate for this  prefactor can have substantial effects on the allowed parameter space for the extended hybrid models that exit inflation through a first-order PT.}. One notices that once the nucleation rate is found, we can determine the probability of PT in the early Universe,
\be
p =\frac{\Gamma}{H^4}\,,
\label{probability}
\ee
where $H$ is the Hubble parameter at the time of PT. Fortunately, for the quartic potential $V_2$ in the second line of (\ref{Potential}), the Euclidean action for a first-order PT has been parametrized extrapolating the  numerical analysis with analytic results \cite{Adams:1993zs},
\begin{equation}
	S_{{}_{\rm E}}= \frac{4 \pi^2}{3 \lambda}(2-\delta)^{-3}(\alpha_1
	\delta+\alpha_2 \delta^2+\alpha_3 \delta^3) \,,
	\label{Euclidian-act}
\end{equation}
where $\alpha_1= 13.832 ,~\alpha_2=-10.819,~\alpha_3=2.0765$, and
$\delta$ is a $\phi$-depended function, which is defined as
\begin{equation}\label{delta}
	\delta=\frac{9 \lambda \alpha}{\gamma^2}+\frac{9 \lambda \lambda'
		\phi^2}{\gamma^2 M^2} \,.
\end{equation}
This relation is only valid in the allowed range, $0 < \delta < 2$, and therefore we should be careful that the legitimate parameter space for PT satisfies this condition. Once $\phi$ meets $\phi_{{}_{\rm I}}$, potential $V_2$ starts to develop another minimum along the $\psi$ direction, which is not energetically favorable for quantum tunneling. This can be observed by noticing that for such values of $\phi$, $\delta$ is bigger than $2$. However, when inflaton rolls down more  along the $\phi$ direction, the other minimum goes down further and finally becomes degenerate with the one at $\psi=0$, at  $\phi_{{}_{\rm crit.}}$. At this point, we have $\delta=2$, and the Euclidean action blows up, leading to $p=0$. When the rolling field drops below its critical value further, $V_2$ at the new minimum turns into a true vacuum, which is when $\delta$ departs from two and moves toward zero such that the probability of PT $p$ becomes more appreciable (see Fig. \ref{psi-direction-potential}). It can be shown that \cite{Guth:1982pn}, the true vacuum percolates effectively provided that $p\gtrsim p_c=0.24$. This is the critical value that we assume for the first-order PT to occur and complete.

Three scenarios are conceivable about how inflation ends and how the field ends up in the true vacuum: In one, the end of inflation takes place when the first slow-roll parameter $\epsilon$ for the inflaton potential $V_1$  becomes one. This is the case when the following relation between the parameters holds:
\begin{equation}\label{epsi-lon}
	\epsilon \equiv \frac{1}{2}
	\left(\frac{V_1^{\prime}}{V_1}\right)^2 = 1\,.
\end{equation}
If we indicate the physical solution of the above polynomial equation as $\phi_{{}_{\rm e}}$, when the inflaton field meets this value, inflation will terminate. In this scenario, the field continues rolling along the ridge, and it could be that
\begin{itemize}
	\item (\textbf{1}) the field $\psi$ settles to the true vacuum due to a tachyonic instability when $M_{{}_{\psi}}^2<0$ ,
	\item (\textbf{2}) the field $\psi$ settles to the true vacuum through a first-order PT after the termination of inflation via slow-roll violation at some $0<\phi_{{}_{\rm pt}}<\phi_{{}_{\rm e}}$ when $p>p_c$ ,
	\item (\textbf{3}) triggering the false vacuum decay occurs before the slow-roll violation occurs, namely inflation abruptly stops via a first-order PT when the value  $\phi_{{}_{\rm pt}}$ is bigger than the slow-roll violating value $\phi_{{}_{\rm e}}$ in the absence of the $\psi$ direction. \footnote{One can also consider a scenario (\textbf{4}), in which inflation and decay of the false vacuum direction happen through the waterfall field. However, this scenario is observationally indistinguishable from the scenario (\textbf{1}), at least from the aspects we assess in this paper.}
\end{itemize}
In all cases, the number of e-folds can be evaluated as
\bea
N(\phi_{{}_{\rm CMB}},\phi_{\rm e})= -\int_{\phi_{{}_{\rm CMB}}}^{\phi_{{}_{\rm e}}}\frac{V_1}{V_1^{\prime}} \, d\phi\,.
\label{e-folds}
\eea

Although this seems a complicated integral, it can be calculated exactly for (\ref{inflection-point}).  Here $\phi_{\rm CMB}$ is the initial value for the inflaton field which can be determined by large scale observations.  Another important parameter  is the scalar spectral index, $n_s$,
\begin{eqnarray}\label{nS}
	n_{s}-1&=&-6 \epsilon + 2 \eta\,,
\end{eqnarray}
where
\begin{equation}\label{eta}
	\eta \equiv  \frac{V''_1}{V_1}\,.
\end{equation}
\begin{figure}[t]
	\includegraphics [width=3in]{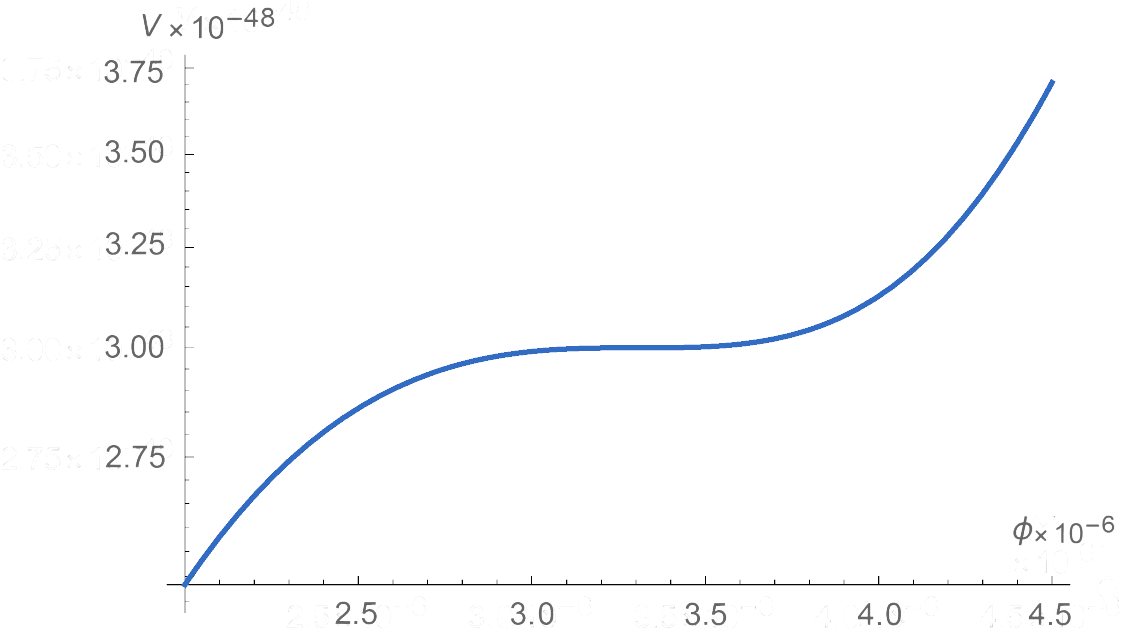}
	\caption{Illustration of inflection-point potential $V_1(\phi)$ with parameters given by (\ref{inf-data}). }
	\label{Fig-v1}
\end{figure}
Moreover, for the inflection-point inflation, $V_1$, the start and the end of the inflationary era happen in a neighborhood of inflection point $\phi_i$.  Making use of (\ref{inflection-point}), one can  expand relations (\ref{epsi-lon})-(\ref{eta}) at the leading order of $\nu$ and drive the following relation for the spectral tilt \cite{Lyth:2006rze}
\begin{equation}\label{ns-beta}
	n_s = 1-4\beta\, \cot\beta N_e\,,
\end{equation}
where $\beta$ is a dimensionless parameter and has been defined as
\begin{equation}\label{bet}
	\beta \equiv \frac{6\nu}{\phi_0^2}\,.	
\end{equation}
In a similar fashion, the power spectrum of fluctuations in terms of $\beta$ and $N_e$ is also given by \footnote{We have used the standard definition of power spectrum $\mathcal{P}_{\phi}=\frac{V_1}{24\pi^2\epsilon}|_{k=aH}$.},
\begin{equation}\label{power}
	\mathcal{P}_{\phi}^{1/2}\approx 0.625\, \frac{m}{\phi_0}\Big(\frac{\sin^2\beta N_e}{\beta^2}\Big)\,.
\end{equation}
Concerning the Planck observation \cite{Akrami:2018odb} for spectral index {\it i.e.} with the mean value $n_s\approx 0.965$ and Eq. (\ref{ns-beta}),  the value of $\beta$ can be specified for a given $N_e$. On the other hand, for the inflection-point inflation the Hubble parameter can be approximated as
\begin{equation}\label{hubble-appro}
	H^2\approx \frac{1}{36}m^2\phi_0^2\,.
\end{equation}
With this estimation and from the  normalization for the amplitude of scalar perturbations, (\ref{power}), the parameters in $V_1$ {\it i.e.} $\phi_0,\, m,\,\nu$. can be completely fixed.

As will be discussed in the next section, the mass of PBHs resulting from the collapsing bubbles or bubble collisions peaks mainly around a value inversely proportional to the Hubble parameter at the formation time. On the other hand, from the recent observational constraints, it is realized that those PBHs with the masses lying within the range of $10^{-17}-10^{-13}$ solar mass can compensate for $10\%$-$100\%$ of the dark matter\cite{Niikura:2017zjd, Montero-Camacho:2019jte}. This would be equivalent to placing the Hubble parameter in the range of
\be
1.3 \times10^{-25}M_{\rm P} \lesssim H \lesssim 1.3\times 10^{-21}M_{\rm P}\,.
\ee
Although, such  $H$'s are very small, it would be always possible to tune the parameters in the inflection-point potential such that all observational criteria are satisfied. For instance, here we study the possibility of PBH creation with $\sim 10^{-14}-10^{-13}$ solar mass. For such a typical mass, the Hubble parameter turns out to be about $H\approx 10^{-24}M_{\rm P}$, which, and assuming an instant reheating, it requires $N_e\approx 40$ to solve the problems of SBB. Accordingly, from Planck observations for the amplitude of scalar fluctuations and the spectral index, which are given by $\mathcal{P}_{\phi}\approx 2\times 10^{-9}$ and $n_s\approx 0.965$ respectively, and also making use of (\ref{ns-beta})-(\ref{hubble-appro}), we tune $V_1$ such that
\bea\label{inf-data}
m&=&1.8048357604\times 10^{-18}\, M_{\rm P} ,\\ \nonumber  \phi_0&=&3.3077384399035\times 10^{-6} \, M_{\rm P} \\ \nonumber \nu&=&5.970452\times 10^{-14}
\eea
The graph of potential $V_1(\phi)$ can be seen in Fig.(\ref{Fig-v1}). For these values of parameters, the points $\phi_{{}_{\rm CMB}}$ and $\phi_{{}_{\rm e}}$\footnote{
	Assuming that inflation ends when $\epsilon=1$.} are given by
\begin{eqnarray}
	\phi_{{}_{\rm e}}&=&3.30566217\times 10^{-6},\, M_{\rm P}\\ \nonumber  \phi_{{}_{\rm CMB}}&=&3.3077384399034\times 10^{-6}\, M_{\rm P}.
\end{eqnarray}

We first focus on the scenario (\textbf{1}) above,  where the field becomes tachyonic after inflation ends via slow-roll violation along $\phi$. As we will see later, the advantage of this scenario is that the small bubbles (subcritical) find enough time to collapse. Moreover, in this case, the large bubbles (supercritical) also find the opportunity to inflate and connect to the parent FRW universe via a wormhole, where the wormhole pinches off with black holes at the mouths. Therefore they can contribute to the PBH mass fraction. We have tuned the parameters in $V_2$ such that during the course of inflation, the probability $p(t)\sim \mathcal{O}(10^{-18})$ and remains almost constant. The direction $\psi$ becomes {\it tachyonic} immediately after the slow-roll violation.  Our parameters are tuned as
\bea\label{tachyonic-data}
\lambda&=&2.6934\times 10^{-10}\,,\, \, \gamma=1.7310\times 10^{-14}\,,\\ \nonumber
\lambda'&=&-\alpha=5.0\times 10^{-28}\,,\, \, M=2.975\times 10^{-6}\, M_{\rm P}.
\eea
For such values of the parameters space, $M_{\psi}^2$ becomes negative when $\phi$ drops below $\phi_{\rm tac}\approx 2.975\times 10^{-6}\, M_{\rm P}$.

\begin{figure}[t]
	\includegraphics [width=3in]{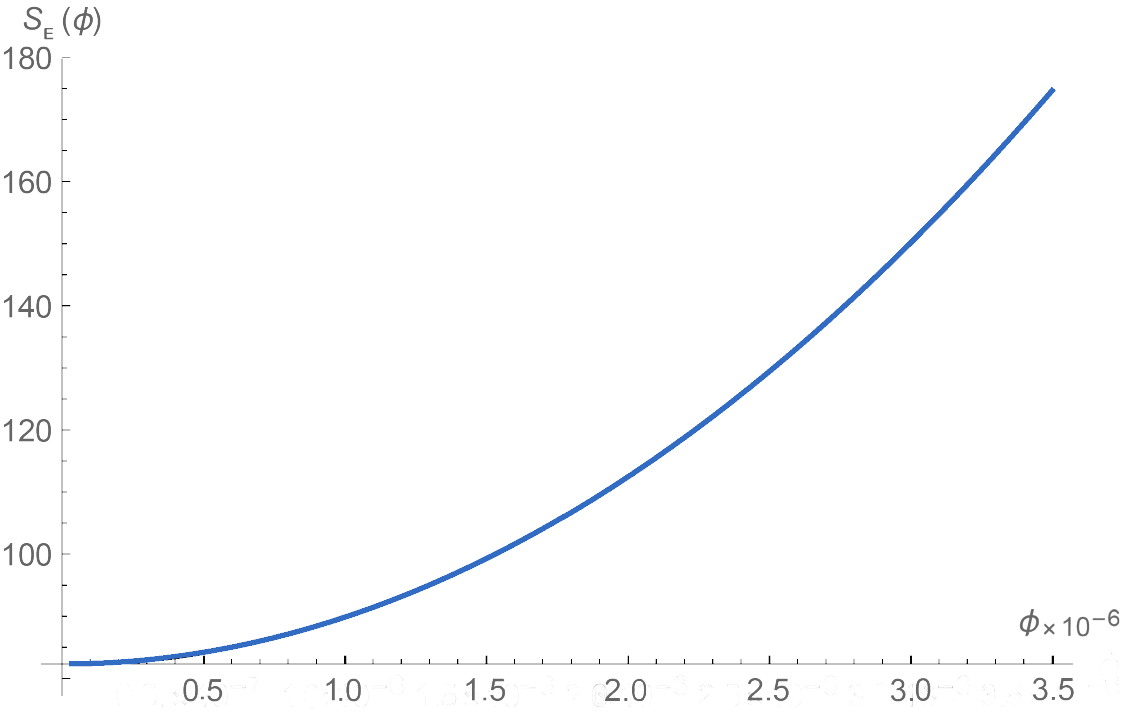}
	\caption{The Euclidean action in terms of $\phi$ . }
	\label{Fig-Se}
\end{figure}

For scenario (\textbf{2}), where the first-order PT kicks in after the end of the inflationary era with the slow-roll violation, we tune other parameters in $V_2$. We assume that the true vacuum percolation happens at $\phi_{{}_{\rm pt}}=10^{-11}\, M_{\rm P}$ for $V_0\approx 3.0\times 10^{-50}\, M_{\rm P}^4$. For that, we choose the following set for the parameters of $V_2$,
\bea\label{data-A}
\lambda&=&1.3011\times 10^{-9}\,,\, \, \gamma=4.7311\times 10^{-14}\,,\\ \nonumber
\lambda'&=&\alpha=5.0\times 10^{-28}\,,\, \, M=3.546946\times 10^{-6}\, M_{\rm P}.
\eea
Inflation ends at $\phi_{{}_{\rm e}}$ via slow-roll violation and the background becomes decelerating since then, until the field rolls to $\phi_{{}_{\rm pt}}$. At this time, the field configuration makes a firs- order PT to the true vacuum. As we will see, this model can produce a gravitational wave (GW) signal from the first-order PT that is observable by LISA. Also, many of the {\it subcritical} bubbles produced during inflation, namely the ones with the radius smaller or about the true vacuum Hubble radius, find enough time to collapse to produce the required PBHs that account for part or all dark matter later. As depicted in  Fig. \ref{Fig-Se}, for these parameters, the Euclidean action in terms of $\phi$ will be decreasing during inflation. Since the inflaton field is decreasing as a function of time, the Euclidean action will be a decreasing function of time too. As soon as the inflaton reaches close to its critical value, $\phi_{{}_{\rm pt}}$ where the probability of bubble nucleation reaches $p_c$, more new bubbles of true vacuum are produced; however, these bubbles do not find time to collapse and coalesce. Instead, they collide to terminate inflation by converting the bubble walls' energy to radiation. In the last section, we will show that the bubble collision can not produce many PBHs to alter the mass distribution of PBHs produced from the collapse of the bubbles. One should also note that the extrapolation of the Euclidean action beyond the time the first-order PT becomes complete does not make sense.
The maximum reheating temperature happens in the case of instant reheating, when the kinetic energy in the bubbles and the inflaton energy are transformed almost completely to radiation. This reheating temperature is given by
\be\label{temperature}
T_{*} = \Big(\frac{90 H_f^2 M_{\rm p}^2}{\pi^2  g_{*}}\Big)^{1/4}\,,
\ee
where $g_{*} \approx 106$, is the total number of relativistic degrees of freedom, and $H_f$ denotes the value of the Hubble parameter calculated
at the end of inflation. For the parameter set in (\ref{inf-data}) , we obtain $T_{*}\approx 5.99\times 10^{-13}\, M_{\rm P}$.

As mentioned before, scenario (\textbf{3}), where inflation terminates via the first-order PT, can happen in this framework too. In the model at hand, like scenario (\textbf{1}), this would not betide unless one assumes $\alpha<0$.  Such a scenario can happen, for example, if we tune the parameters as follows :\footnote{It is important to note that this scenario would not happen if we just allow for a positive value of $\alpha$. In fact, one can simply deduce from relations (\ref{Euclidian-act}) and (\ref{delta}) that if $\alpha$ was a positive number then the transition probability (\ref{probability}) would remain nearly constant while the inflaton is rolling from $\phi_{{}_{\rm CMB}}$ to $\phi_{{}_{\rm e}}$.}
\bea
\lambda&=&8.2345\times 10^{-9}\,, \qquad \gamma=1.9950\times 10^{-13}\,,\\ \nonumber
\lambda'&=&-\alpha=5.0\times 10^{-20}\,, \qquad M=3.305\times 10^{-6}\, M_{\rm P}.
\eea
In the next sections, we first obtain the SGWB resulting from the first-order PT in cases (\textbf{2}) and (\textbf{3}), and then we compute the fraction of PBHs produced during inflation in cases (\textbf{1}) and (\textbf{2}). We also estimate the PBHs produced from bubble collisions in cases (\textbf{2}) and (\textbf{3}) in the last section.

\section{Stochastic gravitational wave background from first-order phase transition\label{SBGWsPT}}

Since some inflationary model with the parameters obtained in the last section, ends with a first-order PT, we first obtain the SGWB signature that these models generate. We employ the gravitational wave spectrum results generated from a first-order PT \cite{Kosowsky:1992vn,Huber:2008hg}. Although the end of inflation happens sometime before first-order PT (and meantime, some reheating may ensue from the coupling of inflaton to other species), still, the real reheating does not begin until the first-order PT completes via bubble collision. This will allow us to use SGWB generation's formalism from a bubble collision at zero temperature because we expect inflation to have supercooled the Universe. During a first-order PT, gravitational waves may be produced by bubble collisions or turbulent plasma motion if the temperature is nonzero. The numerical analysis for two-bubble collisions was first developed in \cite{Kosowsky:1992vn} and extended for more bubbles in \cite{Huber:2008hg}. As discussed there, around the peak frequency, $f_m$, the amplitude of gravitational waves takes the shape of an asymmetric wedge form, which raises in the region with $f<f_m$ as $f^{2.3}$ and declines as  $f^{-1}$ in the region $f>f_m$.

After considering postinflationary redshifts, the today peak frequency $f_m$ can be written in terms of the reheating temperature as \cite{Huber:2008hg}
\be\label{frequency-peak}
f_{\rm m}=3\times 10^{-10} {\left(\frac{g_{\ast}}{100}\right)}^{1/6}  \left(\frac{T_{\ast}}{1  \rm GeV}\right) \left(\frac{\beta}{H_{{}_{\rm pt}}}\right)\,,
\ee
where $H_{{}_{\rm pt}}$ stands for the Hubble parameter at the time of the first-order PT. Moreover, using the Euclidean action, we can also obtain a relation for the amplitude at the above peak frequency. This amplitude is given by
\be\label{GW-amplitude}
\Omega_{{}_{\rm GW}} h^2 (f_m)=10^{-6} \left(\frac{g_{\ast}}{100}\right)^{1/3} {\left(\frac{H_{{}_{\rm pt}}}{\beta}\right)}^2,
\ee
where $\beta$ is
\be\label{beta-time}
\beta\equiv \frac{dS_{{}_{\rm E}}}{dt}= \frac{dS_{{}_{\rm E}}} {d\phi} \frac{d\phi} {dt}.
\ee
The quantity $1/\beta$  provides an estimation of the time span of how long the first-order PT lasts\footnote{To find the velocity of rolling direction in (\ref{beta-time}) at the PT time after the inflation, one has to integrate the field equation $\ddot{\phi} +3H\dot{\phi} + \frac{dV_1}{d\phi}=0$ from the end of inflation to the point of PT, which we did. If the first-order PT occurs right before that inflation ends with slow-roll violation, the velocity is given by $\dot{\phi}^2=\frac{2\epsilon V_1}{3-\epsilon}|_{\phi=\phi_{\rm pt}}$.}. The above analysis for GWs would be reliable if the time duration for which the PT takes to complete is much smaller than the expansion rate, {\it i.e.}, $\beta/H_f  \gg 1$.  For the parameters set in (\ref{inf-data}), the ratio is about $\beta/H_f \sim 16$. This confirms the above discussed numerical tuning and leads to the following values for amplitude and peak frequency
\bea\label{omega-frequency-results-A}
\Omega_{{}_{\rm GW}} h^2 (f_m)&\approx&3.55\times 10^{-9}\,,\\ \nonumber
f_m&=&0.0074\, {\rm Hz}\,.	
\eea
Interestingly, this falls in the LISA sensitivity band,
\begin{figure}[t]
	\includegraphics [width=3in]{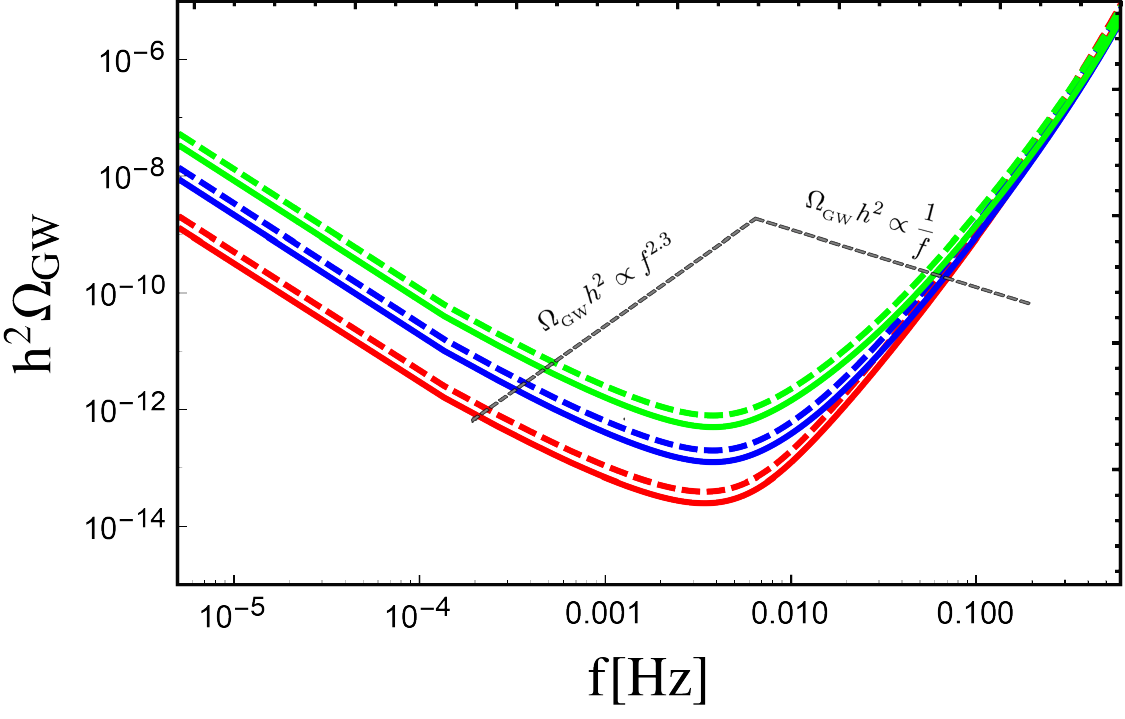}
	\caption{The SGWB signal resulting from the case (\textbf{2}) where inflation ends via slow-roll and first-order PT completes some time after inflation. Both the low and high frequency branches of the produced SGWB spectrum could in principle be observed by LISA. The colored curves depict the power law sensitivity curves for the six LISA configurations from two independent interferometers at low frequencies: green A1M5, green dashed A1M2, blue A2M5, blue dashed A2M2, red A5M5, red dashed A5M2 \cite{Bartolo:2016ami}, where in the notation AiMj i refers to the length of the interferometer arm in millions of Km and j to the duration of the mission in years.}
	\label{Fig-LISA-Case-b}
\end{figure}
and the peak frequency amplitude and the low- and high-frequency branches of the SGWB spectrum fall within the sensitivity band of the  Laser Interferometer Space Antenna (LISA) gravitational wave probe which is aimed to detect the gravitational wave signature in the range $10^{-4}-10^{-1}$ Hz; please see Fig. \ref{Fig-LISA-Case-b}. As we will argue in the next section, some bubbles nucleated during inflation, whose size is very close to the Hubble radius of the true vacuum, find the opportunity to collapse and constitute part of the  dark matter energy density in this example.

The scenario (\textbf{3}) happens if inflation terminates abruptly with the first-order PT before slow-roll violation or tachyonic instability takes place. As mentioned before, in the model at hand, such a scenario would not occur unless one assumes $\alpha<0$. We have to be careful of this assumption because when the value of inflaton decreases below $\phi_{\rm PT}$, the mass $M_{\psi}^2$ also gradually decreases and finally becomes negative. To avoid settling to the true vacuum through rolling, instead of a first-order PT, we must ensure that the time for rolling to the vacuum is much larger than the time it takes for the first-order PT to complete. This is guaranteed if we tune the parameters in $V_2$ such that
\begin{equation}\label{beta-condition}
	\beta \gg |\alpha|^{1/2}M\,.
\end{equation}
This condition guarantees that the timescale for rolling when  $M_{\psi}^2$ becomes negative is much smaller than the time of PT. Such a scenario can happen, for example, if we tune the parameters as follows:
\bea
\lambda&=&8.2345\times 10^{-9}\,, \qquad \gamma=1.9950\times 10^{-13}\,,\\ \nonumber
\lambda'&=&-\alpha=5.0\times 10^{-20}\,, \qquad M=3.305\times 10^{-6}\, M_{\rm P}.
\eea
For these sets of data, the peak frequency of SGWB and its amplitude is
\bea\label{omega-frequency-results-negative}
f_m&=&9.73\times 10^{10}\, {\rm Hz}\,,\\ \nonumber
\Omega_{\rm GW} h^2 (f_m)&=&2.10\times 10^{-35}\,,
\eea
which is completely out of the frequency range and sensitivity of any planned probes. It should come as no surprise because the peak frequency (\ref{frequency-peak}) is proportional to $\beta$ and, in principle, one expects that the condition (\ref{beta-condition}) leads to a large value for peak frequency. The important thing is that, in such cases where inflation terminates with a first-order PT, the production of PBHs only happens through the bubble collision process.  As shown in the next section, this kind of inflation ending never leads to an appreciable number of PBHs.

\section{Mass fraction of PBHs}\label{massfractionpbhs}

In this section, we try to compute if the model with the parameters discussed in the previous section can yield a significant PBH mass fraction compensating for all or part of the missing dark matter. In the context of the bubble nucleation scenario in the early Universe, PBHs can be formed in two ways: the collapsing bubbles and the bubble collision. Below, we will study scenarios related to the models we discussed in the last section.

\subsection{PBHs from collapsing bubbles}

While the field rolls along the $\phi$ direction, inflation proceeds. As soon as the second minimum along the $\psi$ direction forms and becomes degenerate with the one at $\psi=0$, bubbles of true vacuum nucleate and then expand during inflation. Since the nucleation rate enhances very slowly with time in our model, more of such bubbles gradually form during the course of inflation. Here, we simply take the inflationary energy density to be given by $\rho_f$ of the false vacuum,  which changes very slowly during inflation. In the orthogonal direction, the true vacuum also varies with time, and we denote its energy density by $\rho_{t}$. The nucleated bubbles, with the energy density of $\rho_{t}$ inside, initially have a radius smaller than the inflationary Hubble length $H_t^{-1}$, with $H_t= \sqrt{\rho_t/3}$ at the formation time, but they expand very fast and can become large, some even larger than $H_t^{-1}$. This is more probable, particularly for the bubbles that nucleate much earlier than the end of inflation. The tension of  bubbles is related to the Euclidean action at the time of nucleation of the bubble through the relation 
\be
\sigma \simeq  \frac{S_{{}_{\rm E}} H_{\rm t}^{3}}{ 2\pi^2}\,.
\ee
One can define a Hubble parameter related to this tension as
\be
H_{\sigma}=\frac{\sigma}{4M_{\rm P}^2}\,,
\ee
which is the Hubble parameter resulting from the repulsive self-gravitational energy of the bubble wall. In our models, $S_E\sim {\rm few}\times 100$, so $\sigma\sim H_t^3$ and $H_{\sigma}\sim \left(H_t/M_{\rm P}^2\right)^2 H_{t}\sim 10^{-48} H_t\ll H_t$.

We first focus on scenario (\textbf{1}) that we described above, in which inflation ends with the slow-roll violation and settling to the true vacuum occurs with tachyonic rolling. If inflation ends at time $t_e$ due to violation of the slow-roll, which, for example, happened in the first scenario we described above, depending on whether by the end of inflation, the radius of the bubble $R_e$ is bigger or smaller than $H_t^{-1}$, two scenarios are conceivable. For the details of the following estimations for the parameters, we refer the reader to \cite{Garriga:2015fdk}. If $R_e\lesssim H_t^{-1}$, which are dubbed as subcritical, initially the bubble starts accelerating relative to the surrounding matter  and acquires a large Lorentz factor estimated to be about
\be
\gamma_i\simeq \frac{H_f}{H_{\sigma}}\sim 10^{48}\,,
\ee
where the last approximation used the parameters of the scenario (\textbf{1}). Interacting with the matter surrounding the bubble wall, it loses its kinetic energy and comes to rest with respect to the outside FRW universe in the timescale, which is much smaller than $H_{f}^{-1}$. For our case, this timescale is of order \cite{Garriga:2015fdk}
\be
\Delta t\sim \frac{H_{\sigma}}{H_t^2}\sim 10^{-48} H_t^{-1}\,.
\ee
The radius of the bubble grows only by the factor $\frac{\Delta R_e}{R_e}\approx \frac{\sigma H_f}{\rho_t}\approx \frac{H_t^2}{M_{\rm P}^2}\sim 10^{-48}$, before it starts to decouple from the Hubble flow and recede to collapse to form a black hole. Because of the low inflationary scale in our model, the change in the black hole radius will be negligible for subcritical bubbles.
For the subcritical bubbles, one can simply estimate the mass of the resulting PBHs as
\begin{equation} \label{subcritical}
	M_{\rm bh} \simeq \Big(\dfrac{4 \pi}{3} \rho_t + 4 \pi \sigma H_f\Big) R_e^3 = \kappa R_e^3~\,.
\end{equation}
Here, $\sigma$ and $R_e$ are the tension of the bubble wall and the bubble radius at $t\approx t_e$, respectively. The parameter $\kappa$ in (\ref{subcritical}) depends on the Hubble parameter of the true and false vacuum and the tension of the bubble walls. It can be written down as \cite{Deng:2017uwc}
\begin{equation}\label{kap}
	\kappa = \frac{1}{2} H_t^2 + 2H_\sigma \left( \sqrt{H_f^2 - H_t^2} - H_\sigma \right)\,.
\end{equation}
The subcritical bubbles lead to the black holes with mass smaller than the transition mass $M_{\ast}$ defined as \cite{Deng:2017uwc,Garriga:2015fdk}
\begin{equation}
	M_{\ast}\sim\frac{8\pi H_{t}^{3}}{\kappa^{2}}\,.
\end{equation}
Since in our model, $H_t\sim \mathcal{O}(10^{-24}) M_{\rm P}$, we expect that $\kappa \approx \frac{1}{2} H_t^2$ and $M_{\ast}\sim 32\pi M_{\rm P}^2 H^{-1}_t$.

On the other hand, if the bubble size exceeds $\sim \frac{M_{\ast}}{M_{\rm P}^2}$, the bubble continues to inflate.
The inflating baby Universe is connected to the outside FRW universe by a wormhole that eventually pinches off after a timescale $\sim \frac{M_{\rm bh}}{M_{\rm P}^2}\gtrsim 32\pi H_t^{-1}$ and black holes form on two mouths of the wormhole. It can be deduced from causality  that the region influenced by the Schwarzschild radius of the resulting PBHs cannot be larger than the Hubble radius  of the parent Universe, with $t_h=H_f R_e^2$ \footnote{This is  because during the radiation dominated era, the parent Universe with scale factor $a(t)=(t/t_e)^{1/2}$ admits Hubble radius given by $t_h = a(t_h) R_e$, where $R_e$  is the radius of the bubble at the end of inflation.}. This implies that the mass of such PBHs should be of the order of
\be
M_{\rm bh}\simeq \frac{4\pi}{3}\rho(t_h)H_f^{-3}(t_h)=H_f R_e^2\,.
\ee
Nonetheless in both supercritical and subcritical cases, it is supposed that a bubble that nucleates at $t=t_n$ has a radius negligible compared to the Hubble radius $H_t^{-1}$. This easily helps us to approximate the radius of the bubbles of true vacuum as \footnote{Below we will use the fact that in our case to a good approximation, $H_f\approx H_t$ }
\begin{equation}\label{radius}
	R(t)\approx H_f^{-1}\left(e^{H_f (t-t_n)} - 1\right)\, .
\end{equation}
The number of nucleated bubbles in the comoving  four-volume element is
\begin{align}\label{number}
	dN = p(t_n) H_f^4 e^{3H_f t_n} d^3\mathbf{x} dt_n\,.
\end{align}
Using relations (\ref{radius}) and (\ref{number}), we can  determine the number density of bubbles with radius $(R_e , R_e+dR_e)$ during the radiation era as
\begin{equation}\label{population}
	dn(t) = p(t_n) \frac{dR_e}{(R_e + H_f^{-1})^4}\left[\frac{a(t_e)}{a(t)}\right]^3 \,.
\end{equation}
The last cubic factor on the right hand side of (\ref{population}) comes from the fact that for  the black holes formed during the radiation dominated epoch, the corresponding bubble number density should be diluted by the cosmic expansion factor $(a(t_e)/a(t))^3$, where $a(t)\propto t^{1/2}$.  Now, to calculate the black hole mass distribution, we define the standard total fraction of cold dark mater (CDM), which are in the form of PBHs as
\begin{equation}\label{fraction}
	f_{\rm{PBH}}\equiv\frac{\rho_{\rm{PBH}}(t)}{\rho_{\rm{CDM}}(t)}=\int \frac{dM}{M} f(M)\,,
\end{equation}
where $\rho_{\rm{PBH}}(t)$ and $\rho_{\rm{CDM}}(t)$ are the energy density of PBHs and CDM. The CDM energy density during the radiation era changes with time as $\rho_{{}_{\rm{CDM}}}(t)\sim\dfrac{1}{Bt^{3/2}{\cal M}_{{}_{\rm eq}}^{1/2}}$, where $B\sim 10$ and ${\cal M}_{{}_{\rm eq}}\sim 10^{17}M_{\odot}$ is of the order of the CDM mass in the Hubble radius at the equality time. Above, $f(M)$ is called the standard  mass function, which is defined as
\begin{equation}\label{mass-fun}
	f(M)=\frac{M^{2}}{\rho_{{}_{\rm{CDM}}}(t)}\frac{dn(t)}{dM}.
\end{equation}
It can be shown that, the while mass function of PBHs of subcritical mass remains constant, it decreases with $M$ for the supercritical cases. In a more exact form, \cite{Deng:2017uwc}
\begin{equation}\label{approx-mass-fun}
	f(M)\sim B p(t_n){\cal M}_{{\rm eq}}^{1/2}\begin{cases}
		M_{\ast}^{-1/2} & M<M_{\ast}\\
		M^{-1/2} & M>M_{\ast}\,,
	\end{cases},
\end{equation}
\begin{figure}[t]
	\includegraphics [width=3in]{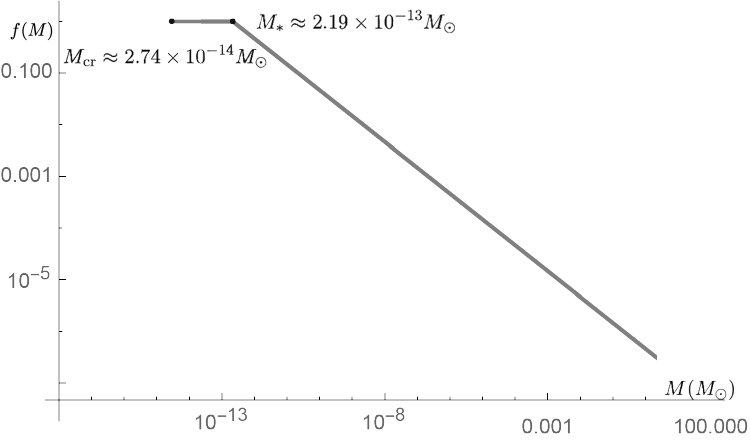}
	\caption{The mass distribution of PBHs formed from the bubbles nucleated during the double-field inflation in scenario (\textbf{1}). The mass distribution for subcritical bubbles is almost constant for $M_{{}_{\rm cr}}\lesssim M \lesssim M_{\ast} $, but it drops like $M^{-1/2}$ for $M \gtrsim M_{\ast}$.}
	\label{PBHs-mass-distribution-a}
\end{figure}
One may realize a lower bound mass for the subcritical PBHs due to the shape fluctuations of the bubbles. In fact, quantum fluctuations may cause the nucleated bubbles of true vacuum not to be exactly spherical at the time of nucleation \cite{Garriga:1991tb}.  With the collapse of a subcritical bubble, the shape fluctuations increase and might finally be large enough before the bubble becomes a PBH. In such case, the bubble has a different fate: instead of ending up as a black hole, it will be  shattered into relativistic particles. This sets a lower bound for the mass of the subcritical bubbles as follows: \cite{Deng:2017uwc}
\begin{equation}
	M_{\rm min}\sim \begin{cases}
		{\rm max} \left\{M_H, M_F\right\}, & M_F<M_{\rm cr}\\
		M_{\rm cr}, & M_F>M_{\rm cr}
	\end{cases},
\end{equation}
where
\begin{eqnarray}
	M_F &\equiv& \rho_{t}\left(\frac{\rho_f }{\rho_{t} \sigma}\right)^{3/2},\\ \nonumber
	M_{H} &\equiv& \kappa H_{f}^{-3} ,\\ \nonumber
	M_{\rm{cr}}&\sim&4\pi \min\{H_{t}^{-1},H_{\sigma}^{-1}\}.
\end{eqnarray}
In our case, since $\rho_t \approx \rho_f = 3 H_f^2$, it is easy to check that $M_{\rm min}\sim M_{\rm cr}\sim 4\pi M_{\rm {p}}^2 H_f^{-1}$. Therefore the subcritical PBHs lie within the narrow mass band $4\pi M_{\rm p}^2 H_f^{-1}\lesssim M_{\rm sub} \lesssim 32\pi M_{\rm p}^2 H_f^{-1}$. On the other hand, from (\ref{fraction}) and (\ref{approx-mass-fun}) one can find a simple relation for total mass fraction as
\begin{equation}
	f_{{\rm PBH}}\sim B p(t_n)\left(\frac{{\cal M}_{{\rm eq}}}{M_{*}}\right)^{1/2}\left[\ln\left(\frac{M_{*}}{M_{{\rm min}}}\right)+1\right], \label{approxfPBH}
\end{equation}
where the first logarithmic term counts for subcritical bubbles contribution and the second term in the bracket comes from integration over supercritical ones. It is again easy to see that since during the course of inflation $p(t) \sim 10^{-18}-10^{-17}$  for the parameter set (\ref{data-A}), the total mass fraction of CDM becomes $f_{{\rm PBH}}\sim 1$. We have depicted the mass distribution of PBHs resulting from scenario (\textbf{1}); please see Fig.(\ref{PBHs-mass-distribution-a}).

In scenario (\textbf{2}), where the first-order PT occurs at time  $\Delta t\sim 10^{-6} H_f^{-1}$ after the end of inflation, and typical timescale of phase transition is $\sim \frac{H_f^{-1}}{16}$, supercritical bubbles that need about $\sim H_f^{-1}$ to form black holes,  will collide with each other and with other bubbles (supercritical or subcritical) and get destroyed. However, much of the subcritical ones will find the opportunity to collapse in the meantime, in particular, that the ones contributing to the mass function are those whose radius is very close to $H_f^{-1}$. In this case,  the mass fraction of the formed subcritical PBHs is \footnote{This is because the logarithm in \eqref{approxfPBH} is $\sim 1$ and therefore the order of the magnitude of mass fraction is mainly given by $f_{{\rm PBH}}\sim B p(t_n)\left(\frac{{\cal M}_{{\rm eq}}}{M_{*}}\right)^{1/2}$}
\begin{equation*}
	f_{{\rm PBH}}\sim B p(t_n)\left(\frac{{\cal M}_{{\rm eq}}}{M_{*}}\right)^{1/2}\ln\left(\frac{M_{*}}{M_{{\rm min}}}\right)\, \sim \, 1.
\end{equation*}

In scenario (\textbf{3}), where inflation ends and settles in the true vacuum with a first-order phase transition in a timescale $\beta^{-1}\sim 10^{-9}H_f^{-1}$, we do not expect that the nucleated bubbles, either critical or subcritical, will have time to form black holes and all get destroyed and their wall energy is converted to radiation. Both in scenarios (\textbf{2}) and (\textbf{3}), there is a chance that bubble collision leads to concentration of the energy and the formation of the bubbles, but as we will show in the next subsection, the contribution of PBHs from this process is small.

\subsection{PBHs from bubble collisions}

In case of bubble collision giving rise to the PBHs formation, we need to know the probability that $n$ bubbles collide at a point in space-time. According to \cite{Hawking:1982}, the probability that no bubble has
nucleated in the past at a point {\bf p} in space-time is approximately given by
\begin{equation}
	P_0 = e^{-\frac{4\pi}{3}p(t_n)H(t_{\rm p}-t_n)}\,.
\end{equation}
Imagine a ball {\bf U} of radius $r < H^{-1}$ in the neighborhood of {\bf p}. If {\bf O} is the region to the past of {\bf U}
such that it does not coincide with the past of {\bf p}, then its volume turns out to be of the order of $\frac{4\pi}{3} r H^{-3}$.
If there are  $n$ bubbles that have been nucleated in the region {\bf O}, the energy of their walls confined within the ball {\bf U} is approximately  $E_{\rm wall} \approx n\pi r^2 H M^2_{\rm P}$.
This consideration that the energy that gives rise to the gravitational collapse if it is bigger than $4\pi r M^2_{\rm P}$, limits the number of needed bubbles as $n > \frac{4}{r H}$.

However, the above criterion for gravitational collapse holds only for regions whose radius $r$ is
small compared to the Hubble radius, i.e., $H^{-1}$, and hence we can neglect the effects of the curvature and expansion. It is reasonable to assume $r \lesssim \frac{1}{2}H^{-1}$ and thus $n\geq 8$.

Therefore, at least eight bubbles should collide in region {\bf U} to produce a PBH whose mass is $\dfrac{2\pi}{H}M^2_{\rm P}$. The probability of the collision of eight  bubble walls within the ball with $r=\frac{1}{2}H^{-1}$ is simply the product of the probability $P_0$ that bubbles do not nucleate in the past of {\bf p} with the probability of nucleation of eight bubbles in the region {\bf O}. This probability is of  order of  \cite{Hawking:1982}\footnote{This is the way Hawking  {\it et al.} simply found the probability of PBHs formation from bubble collision scenario. Although, they did not provide any estimation of corresponding mass function in their own original paper.}
\begin{equation}\label{prob-bubble}
	P_{\rm col}\,  \sim\,  e^{-\frac{4\pi}{3}p(t_n)H(t_{\rm p}-t_n)}\left(\dfrac{2\pi}{3}p(t_n)\right)^8 \, .
\end{equation}

To find the mass fraction $f(M)$ for the PBHs from collision, first we simply write down the number density of bubbles with radius  $(R_e, R_e + dR_e)$ at the end of inflation  $t_e$ as
\be \label{number-density2}
dn(t_e)=\frac{dN}{dV}=\frac{p(t_n) a(t_n)^3 d^3X dt_n}{a(t_e)^3 d^3X}.
\ee
Using this number density one can obtain the mass fraction  normalized to the dark matter (DM) density:
\begin{equation}
	f(M) = \dfrac{M_{PBH}}{\rho_{\rm DM}(t_e)} \int_{t_i}^{t_p} \frac{dN}{dV} P_{\rm col} dt_n\,,
\end{equation}
where $t_{p}$ is the time at which the first-order PT completes and	 $t_i$ is the beginning of PT.
To find the order of magnitude of mass fraction $f(M)$ for  PBHs formed from bubble collisions, it is enough to use relations (\ref{prob-bubble}) and (\ref{number-density2}) and calculate the following integral for the given model
\bea\label{mass-fraction-collision}
f(M)\, &\sim&\, \Big(\dfrac{2\pi}{3}\Big)^9\dfrac{3 M^2_{\rm P}}{H(t_{\rm p})\rho_{\rm DM}}\\ \nonumber &\times&\int_{t_i}^{t_{\rm p}} p^9(t') H^4(t') e^{-\frac{4\pi}{3}p(t')H(t_{\rm p}-t')}   dt'\, .
\eea

If we use the set of parameters in (\ref{data-A}), the model predicts PBHs of bubble collision provide us only $f(10^{-14}M_{\odot})\sim 10^{-50}$.
Hence, we conclude that due to the first-order PT time being much shorter than the Hubble time, $\beta/H \gg 1$, and the low-energy scale of $H$, the  PBHs coming from the bubble collision cannot compose for a considerable ratio of dark matter. This does not much change even in the case of scenario (\textbf{3}), where $p(t)$ increases up to $0.24$ at the end of inflation. The reason is that  the mass function in (\ref{mass-fraction-collision})  is suppressed by a factor $(H(t_{\rm p})/M_{\rm P})^3$.

\section{Concluding remarks}

Among a host of  PBH formation scenarios, black holes may also form during inflation through the nucleations of bubbles of true vacuum, with a cosmological constant smaller than the one at the false vacuum that drives inflation. This scenario is quite probable in the string theory landscape, where there would be a lot of rolling and tunneling directions simultaneously around. Under the assumption that inflation has happened in such a landscape and enough PBHs as dark matter have been produced during inflation from nucleating bubbles, we constructed a realistic double-field model similar to extended hybrid inflation in which the rolling direction is replaced with a near-inflection-point inflation potential. The model is capable of producing observables compatible with the latest CMB data and producing enough PBHs in the mass range $10^{-17} \lesssim M_{\rm PBH}\lesssim 10^{-13}~M_{\odot}$, where PBHs can constitute up to almost all dark matter energy density. This constrains the scale of inflation to be in the range $10^{-7} \lesssim H \lesssim 10^{-3} ~{\rm GeV}$, which makes the possibility of detection of B modes at the CMB scales fruitless.

During inflation, due to the existence of another direction with smaller vacuum energy, bubbles of true vacuum with small radii, often smaller than the Hubble radius in the true vacuum  $32\pi H_t^{-1}$, form and then stretch to much larger lengths due to the inflationary dynamics. Some even become larger than $32\pi H_t^{-1}$ and start inflating again due to the vacuum energy inside the bubble. The resulting inflating baby Universe is connected to the parent radiation dominated FRW universe after inflation through a wormhole, which pinches off on the timescales of $t_{\rm pinch}\simeq \frac{M_{\rm bh}}{M_{\rm P}^2}$. These bubbles are called supercritical. For our specific inflationary model it turned out that $t_{\rm pinch}\simeq H_f^{-1}$. On the other hand, bubbles that remain smaller than $32\pi H_t^{-1}$ lose their energy on a very short timescale and start collapsing to black holes very fast.

Therefore, how inflation in such a model ends or how the transition from the false to the true vacuum occurs, has consequences for what sections of the mass fraction of the black holes survive the postinflationary phase. In one of the scenarios examined in this paper, inflation ended with a first-order PT. Also, the timescale for the first-order PT to complete is much smaller than the Hubble time during such a PT. Since all such the sub- and supercritical bubbles produced during the course of inflation will collide and get destroyed to convert the energy difference between two vacua to radiation, it is expected that almost all these bubbles will not find enough time to collapse, in the case of subcritical bubbles, or form during the wormhole pinch-off, in the case of supercritical ones. Also, since the timescale of PT will be quite short, the gravitational spectrum from the collision of the bubbles is shifted to very high frequencies and small intensities. This will make the detection of such a SGWB very difficult. In principle, though it should be possible to design a scenario where the first-order phase transition from the false valley to the true vacuum occurs in a reasonable timescale such that the peak frequency of the SGWB is not shifted to such high frequencies. So the absence of detection of PBHs in both branches of the mass distribution, even if a SGWB typical of first-order phase transition is detected, points toward an inflationary scenario in the landscape in which the exit from inflation and the false vacuum valley happens through a first-order phase transition.

On the other hand, if inflation ends via a tachyonic instability, as in hybrid inflation \cite{Linde:1993cn}, both the sub- and supercritical branches of the mass distribution of PBHs survive. There will be no SGWB from a first-order PT in these models, although it could be that the coupling of the inflaton after it rolls toward its true minimum gives rise to SGWB from the usual preheating mechanism \cite{Khlebnikov:1997di, Dufaux:2007pt, Easther:2006gt, Ashoorioon:2013oha}. However, the characteristic of such SGWB spectrum is often different from the ones generated from the bubble collisions. In particular, the rise and fall off in, respectively, the high- and low-frequency branches of the spectrum is different. Also, the SGWB spectrum resulting from preheating often shows a double bump feature, in contrast to the single bump feature of the SGWB generated from a first-order phase transition. Hence, we conclude that detection of PBHs in both branches of the mass distribution is typical of inflationary models in which the exit from the false vacuum valley to the true one occurs through a second-order phase transition.

There is an intermediate scenario in which both these two signatures from inflation within the landscape remain. That is when the first-order PT happens sometime after the end of inflation via slow-roll violation, and the first-order phase transition timescale is not much smaller than the inflationary Hubble time. In this case, much of the subcritical bubbles find enough time to collapse and contribute to the PBH mass fraction. We tuned the parameters in this case such that we would get the whole dark matter energy density from PBHs. From the collision of the bubbles, we also obtained a SGWB signal that falls both in terms of frequency and intensity in the LISA sensitivity band. The energy scale of inflation in our case is typical of a SGWB signal that falls in the LISA sensitivity band, if the timescale of PT, parametrized in terms of $\beta$ takes reasonable values, {\it i.e.} ${\rm few}\lesssim \frac{\beta}{H}\lesssim 50$.  Faster PT will shift the maximum frequency to higher values and suppress the SGWB amplitude. Hence the detection of SGWB spectrum typical of first-order phase transition, and only the subcritical branch of the PBHs mass distribution is indicative of a scenario in which inflation ends with violation of the slow-roll, but settling to the true vacuum occurs via a first-order phase transition.

It is also worthwhile that one investigates the probability of each of the above scenarios in the full parameter space of the model. In fact, in the model at hand, due to the low-energy scale of rolling direction, one naturally expects tuning of parameters. However, this feature of inflection-point inflation is compensated noting that in a potential landscape, inflection inflation is an attractor \cite{Allahverdi:2008bt}. Another level of tuning occurs when we want to  choose the parameters of the rolling inflection-point potential to make it compatible with the CMB observations. This will uniquely determine the parameters in the rolling part of the potential, if the Hubble parameter is determined by the mass of PBHs. However to achieve different exit scenarios, one has to tune the space of parameters in the interacting part of the potential. Of course, since several parameters are involved in this part of the potential, the specification of parameters is not unique and many different sets of parameters can yield similar observables. Sampling the parameter space using the Monte Carlo method can determine the level of fine-tuning required to realize different scenarios. This goes beyond our attempt in this paper, but we plan to refer to it in future.

It seems that, under the assumption that inflation has happened in the string theory landscape and PBHs constitute a sizable portion of dark matter energy density, we can gain invaluable information about the structure of the landscape around us. Although here we focused mainly on the PBHs mass window, $10^{-17}~M_{\odot}\lesssim M_{\rm PBH}\lesssim 10^{-13}~M_{\odot}$, one can design a similar double-field potential at other energy scales. For example, this could have been done for the $10^{-22} \lesssim H \lesssim 10^{-20} ~{\rm GeV}$ inflationary model. The reason for the lower bound of $H\simeq 10^{-22}~{\rm GeV}$ is that we have demanded the reheating temperature resulting from the PT to be bigger than $10^{-2}$ GeV needed for BBN. The range of PBHs produced from such an inflationary scenario is  $4 M_{\odot}\lesssim M_{\rm bh} \lesssim 3650  M_{\odot}$. Interestingly this covers the gap produced by (pulsational) pair-instability supernova processes too \cite{Abbott:2020tfl}. Hence, inflation at such a low scale within the landscape can account for PBHs observed by LIGO, including the ones in the mass gap too. Interestingly the signal from such an accelerated phase of expansion within the double-field model we described can account for the 12.5-yr signal observed by the NANOgrav probe \cite{Arzoumanian:2020vkk}. We plan to return to this in more detail in future work.

\bibliography{bibtex}

\end{document}